\documentclass[aps,twocolumn,prb,10pt,floatfix]{revtex4-2}

\setcitestyle{super,open={},close={}}
\usepackage{pdfcomment} 	
\usepackage[UKenglish]{isodate} 

\usepackage{amsmath} 		
\usepackage{amssymb} 		
\usepackage{chemscheme}     
\usepackage{chemformula}    

\usepackage{bm} 			
\usepackage{breqn} 			

\usepackage{booktabs}	    
\usepackage{multirow}	    
\usepackage[flushleft]{threeparttable}
\usepackage{floatrow}       
\usepackage{color}
	\definecolor{mygray}{rgb}{0.8, 0.8, 0.8}
	\definecolor{coolblack}{rgb}{0.0, 0.18, 0.39}

\usepackage{lipsum}         
\usepackage{hyperref}       
	\hypersetup{colorlinks=true,allcolors=coolblack}

\usepackage{epstopdf} 					
\def\bibsection{\section*{\refname}} 	

\begin{document}
\date{\today}

\author{Philipp J. Heckmeier*, Jeannette Ruf, Brankica G. Jankovi\'{c}, and Peter Hamm\\\textit{Department of Chemistry, University of Zurich, Zurich, Switzerland}\\  *philipp.heckmeier@chem.uzh.ch}

\title {MCL-1 promiscuity and the structural resilience of its binding partners}

\begin{abstract}
{MCL-1 and its natural inhibitors, the BH3-only proteins PUMA, BIM, and NOXA regulate apoptosis by interacting promiscuously within an entangled binding network. Little is known about the transient processes and dynamic conformational fluctuations that are the basis for the formation and stability of the MCL-1/BH3-only complex. In this study, we designed photoswitchable versions of MCL-1/PUMA and MCL-1/NOXA, and investigated the protein response after an ultrafast photo-perturbation with transient infrared spectroscopy. We observed partial $\alpha$-helical unfolding in all cases, albeit on strongly varying timescales (1.6~ns for PUMA, 9.7~ns for the previously studied BIM, and 85~ns for NOXA). These differences are interpreted as a BH3-only-specific ``structural resilience'' to defy the perturbation while remaining in MCL-1's binding pocket. Thus, the presented insights could help to better understand the differences between PUMA, BIM, and NOXA, the promiscuity of MCL-1 in general, and the role of the proteins in the apoptotic network.}
\end{abstract}
\maketitle

Protein-protein interactions are the fundamental driving force for a majority of cellular processes\cite{Jeong2001,Hartwell1999}. Understanding the molecular mechanisms behind this protein-protein interplay is of highest scientific interest\cite{Jin2005}. For numerous protein complexes, it is not clear how they form, how small conformation fluctuations contribute to the complex function and stability, and whether there are intertwined intermediate states of altered conformation \cite{Tsai1999a,Sugase2007}. Illuminating the nuances of these dynamic processes is particularly essential for complexes formed by intrinsically disordered proteins\cite{Wright2009}. These proteins do not assume an ordered structure in their isolated form but only when they are bound to their complex partner. For intrinsically disordered proteins, the process of complex formation can be explained with models\cite{Kumar2000,Boehr2009} such as the induced fit model\cite{Koshland1958}, the conformational selection model\cite{Tsai1999a}, or a hybrid version of both theories\cite{Csermely2010}. 
Intrinsic disorder plays a significant role in promiscuous protein networks, as it enables the complex formation with numerous binding partners\cite{Dunker2005,Kim2008}. Being on ``the edge of chaos''\cite{Uversky2013} ensures structural and functional flexibility and provides an ideal basis for diverse protein-protein interactions, for instance in a network of activator, inhibitor, and effector proteins.

\begin{figure}[t]
	\centering
	\includegraphics[clip, trim=0.3cm 0cm 0.3cm 0cm, width=1\textwidth]{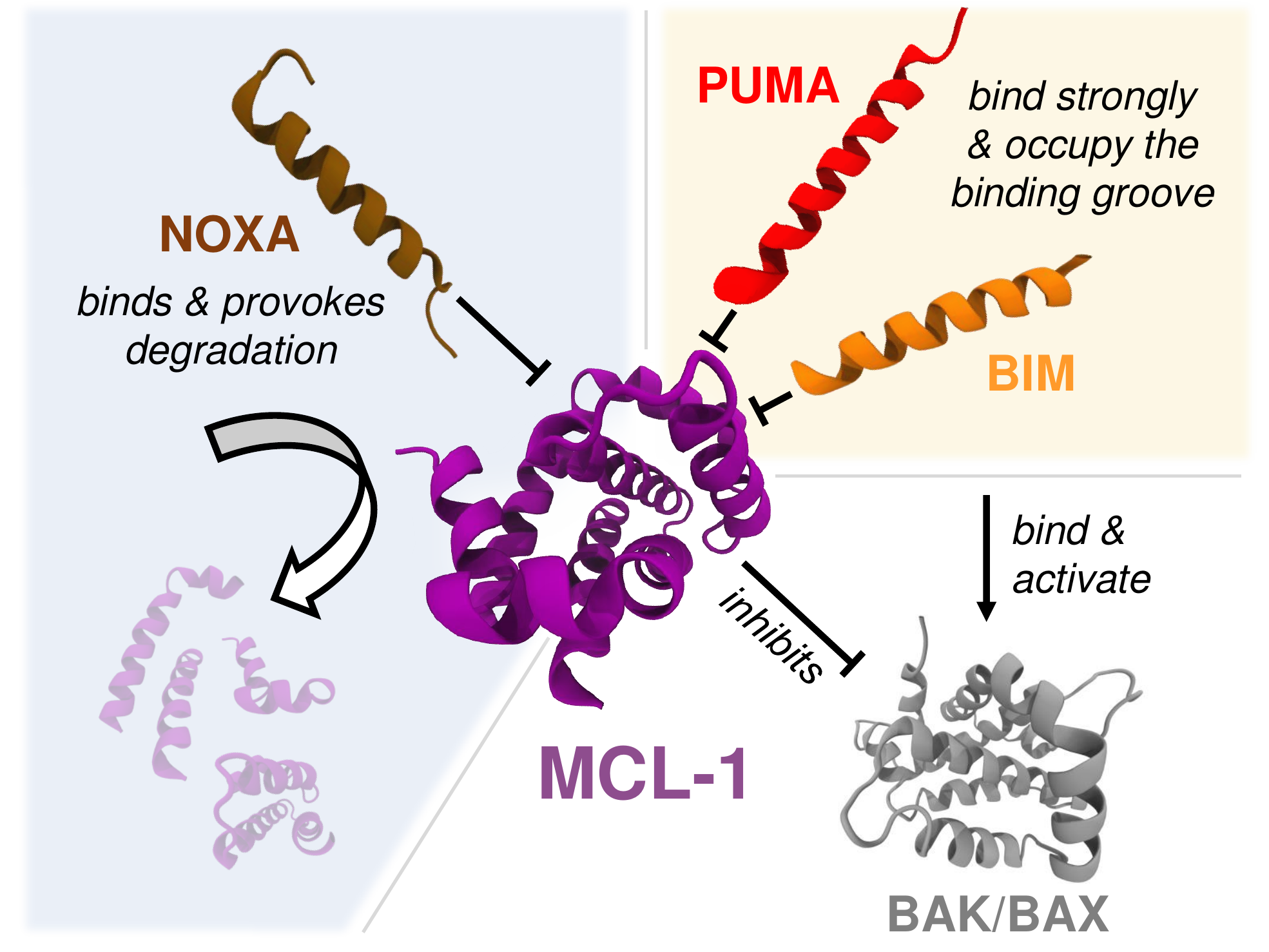}
	\caption{The protein MCL-1 promiscuously binds BH3-only peptides NOXA, PUMA, and BIM, and has a central position in an entangled binding network, regulating apoptotic effector proteins BAK and BAX.}
	\label{fig:Introduction}
\end{figure}

The BCL-2 protein family is a paramount example for an intricate protein network, which is driven by promiscuous interactions of several intrinsically disordered protein domains. In this protein family, categorized in subfamilies based on the type and the number of their BCL-2 homology (BH) domains, the disordered binding domains of so-called BH3-only proteins -- they solely have a BH3 domain -- form complexes with numerous other BCL-2 proteins, thereby controlling apoptosis in a finely-balanced manner\cite{Youle2008,Kale2018,Hinds2007,Dahal2018}. The BH3-only proteins, abundant when cells suffer cytotoxic stress\cite{Adams2007,Czabotar2014,Roufayel2022}, either directly activate the pro-apoptotic effector proteins (BAK and BAX) or inhibit the anti-apoptotic effector-inhibitors such as the Myeloid Cell Leukemia~1 protein (MCL\mbox{-}1), one of the key players in apoptosis regulation\cite{Kale2018} (Fig.~\ref{fig:Introduction}). MCL-1 is overexpressed in various tumor variants, which makes it of high interest in therapeutic application \cite{Bolomsky2020,Adams2007,Thomas2010,Craig2002,Fletcher2019,Bonneaud2022}. At its canonical binding site, MCL-1 promiscuously binds the $\alpha$-helical binding domain of the BH3-only proteins PUMA, BIM, and NOXA\cite{Czabotar2007,Day2008}, most likely by induced fit\cite{Miles2016,Rogers2013,Rogers2014a,Rogers2014b,Heckmeier2022}. PUMA and BIM inhibit MCL-1 by occupying its binding pocket with affinities in the sub-nanomolar range, but additionally bind other anti-apoptotic factors and the pro-apoptotic effector proteins BAK and BAX\cite{Dahal2018,Mei2005,Czabotar2007}. In contrast, NOXA binds MCL-1 specifically and with a weaker affinity, however effects the degradation of the whole complex\cite{Willis2005}.

In their intriguing review\cite{Kale2018}, Kale, Osterlund, and Andrews describe this interplay of different pro- and anti-apoptotic factors very figuratively as a ``dance'' of various partners -- an interesting and vivid depiction of protein promiscuity. In this ``apoptotic dance'', the protein affinity and stability determine the complex formation and whether cell death will be initiated or not. 

To date, little is known about the protein dynamics behind the MCL-1/BH3-only complex stability and formation, let alone intermediate states of BH3-only folding and unfolding at MCL-1's binding groove\cite{Sora2022}.
Investigating subtle, dynamical rearrangements inside proteins like the MCL-1/BH3-only complex is connected to a fundamental challenge: the ability to resolve structural flexibility and small conformational fluctuations in a reasonable time frame\cite{Kumar1999,Tsai1999b}. In the past, the dynamics of protein ensembles -- intrinsically disordered or folded -- have been experimentally studied via single-molecule FRET spectroscopy \cite{Aviram2018,Borgia2018}, NMR spectroscopy \cite{Kovermann2017,Lange2008}, as well as transient infrared (IR) spectroscopy \cite{Buchli2013,Jankovic2021}. IR spectroscopy allows the differentiation of very small conformational differences\cite{Barth2007} and, in its transient form, the sensitive detection of non-equilibrium processes\cite{Lorenz-Fonfria2020}, making it a highly suitable method to study MCL-1/BH3-only complexes.

To selectively trigger a dynamical process inside a protein for transient IR spectroscopy, a fast and precisely induced perturbation of the proteins is required, ideally initiated by short light pulses. In this regard, a plethora of photoreceptor proteins, i.e., light-sensitive or fluorescent, have been investigated in the past \cite{Gerwert1990,Buhrke2020,Buhrke2022,Lorenz-Fonfria2013,Laptenok2018,Kennis2007,Kottke2017}. Beyond proteins that show natural photo-activity, linking azobenzene photoswitches covalently to selected protein domains -- most prominently $\alpha$-helical structures -- defines a potent strategy to introduce photo-sensitivity in molecules, which are otherwise ``blind'' to  light. The light-induced isomerization  of the cross-linked azobenzene moiety leads  to a fast perturbation of the secondary structure\cite{Kumita2000}, and in turn to a slower protein response, both of which can be detected via transient IR spectroscopy \cite{Ihalainen2007,Wachtveitl2004,Rampp2018}. With this technique, the allosteric signal propagation in PDZ domains and the unbinding in the RNase S complex were investigated \cite{Buchli2013,Bozovic2020b,Jankovic2021}. Similarly, we recently revealed the signal propagation inside the MCL-1/BIM complex.\cite{Heckmeier2022}

\begin{figure}[t]
	\centering
	\includegraphics[clip, trim=0cm 0cm 0cm 0cm, width=1\textwidth]{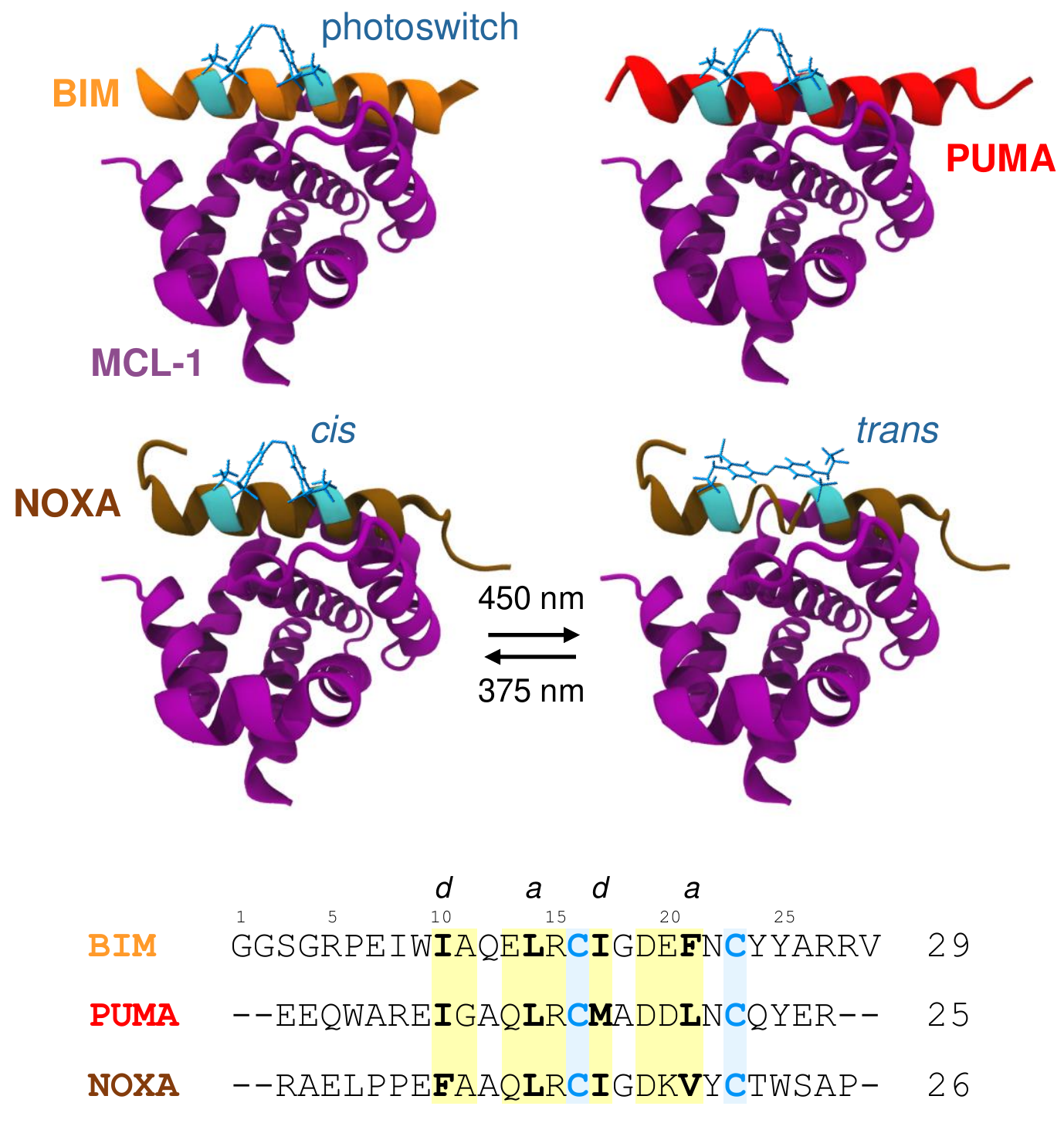}
	\caption{Structures of MCL-1 (purple) binding BIM (orange), PUMA (red), or NOXA (brown), adapted from PDB entries 2NL9\cite{Czabotar2007}, 2ROC\cite{Day2008} and  2ROD\cite{Day2008}, respectively. The peptides were covalently crosslinked with an azobenzene photoswitch (blue) at introduced Cys residues (cyan). Illumination with 450 nm light promotes the isomerization of the photoswitch from the \textit{cis}- to the \textit{trans}-state, exemplified here for NOXA. The opposite direction can be induced with 375 nm light. The $\alpha$-helical BH3-only peptides are destabilized in the \textit{trans}-state. Aligned peptide sequences are also shown together with the \textit{a} and \textit{d} positions of their heptad pattern. The residue positions that are known to form the contact interface with the binding groove of MCL-1 are marked in yellow \cite{London2012}.} \label{fig:Peptides}
\end{figure}

In this study, we apply transient IR spectroscopy to explore the protein dynamics of the intrinsically disordered binding domains of PUMA and NOXA in complex with MCL-1. An azobenzene photoswitch has been covalently linked to the short PUMA and NOXA peptides. The induced isomerization of the photoswitch leads to a subsequent destabilization of the secondary structure of the linked peptides, yet remaining bound to MCL-1 (Fig.~\ref{fig:Peptides}). 
Together with data from the previously analyzed MCL-1/BIM complex\cite{Heckmeier2022}, we classify the kinetic response of these three MCL-1/BH3-only complexes and discuss how the observed protein dynamics integrate into the promiscuous nature of the MCL-1/BH3-only system.

We designed photoswitchable variants of the bindings domains of PUMA and NOXA in complex with MCL-1, to compare them to the previously generated photoswitchable MCL-1/BIM variant \cite{Heckmeier2022}. The domains, from here on \textit{pars-pro-toto} referred to as PUMA, NOXA and BIM, are 25 to 29 amino acid long intrinsically disordered peptides that become $\alpha$-helical when complexing with MCL-1 \cite{Hinds2007}. They interact with a classical heptad pattern with hydrophobic side chains at the a/d positions of the helix (Fig.~\ref{fig:Peptides}, sequences). On the solvent-exposed side of PUMA and NOXA, we introduced two cysteine residues, which were used to covalently bind the photoswitch 3,3'-bis(sulfonato)-4,4'-bis-(chloroacetamido) azobenzene (BSBCA)\cite{Zhang2003} to the peptide. In a previous study with BIM, we identified positions 16 and 23 (two consecutive c positions of the heptad pattern) as anchoring points for the azobenzene moiety. For the newly generated complexes, we introduced cysteines at the corresponding positions (Fig.~\ref{fig:Peptides}, in cyan). 

\begin{figure}[t]
	\centering
	\includegraphics[clip, trim=1.2cm 0cm 0.5cm 0cm, width=1\textwidth]{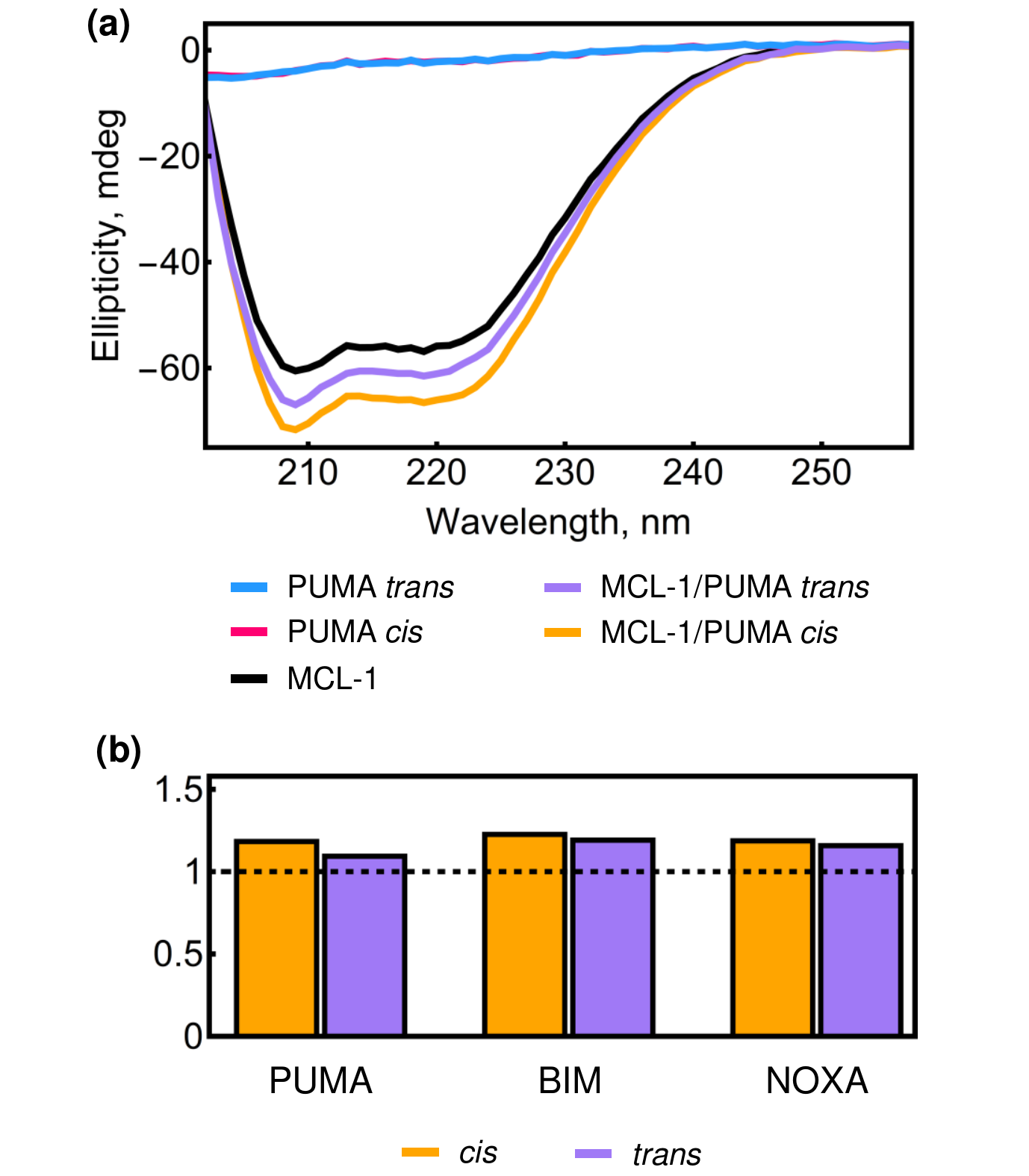}
	\caption{(a) CD spectra of MCL-1 and PUMA (both 20~µM) in isolation and when forming a complex. The PUMA spectra are shown for both the \textit{cis}- and the \textit{trans}-state of the photoswitch. (b) Ellipticity at 220~nm of the MCL-1/peptide complexes relative to that of MCL-1 without peptide (dashed line). The data for BIM are adapted from Heckmeier et al.\cite{Heckmeier2022}.} \label{fig:CDspec}
\end{figure}

Circular Dichroism (CD) spectroscopy of the isolated BH3-only peptides reveal a  random coil structure in both the \textit{cis} and the \textit{trans}-state of the photoswitch (exemplified for PUMA in Fig.~\ref{fig:CDspec}a, red and blue). The CD spectrum of MCL-1 alone displays the classical response for an $\alpha$-helical structure (Fig.~\ref{fig:CDspec}a, black). The $\alpha$-helical content is increased by a factor of $\approx$1.2 when the BH3-only peptides BIM, PUMA, and NOXA are added in equivalent amounts in the dark, in which case the photoswitch is in the \textit{trans}-state (Fig.~\ref{fig:CDspec}a, purple). This proves that the photoswitchable BH3-only peptides assume an $\alpha$-helical structure in the presence of their natural binding partner MCL-1. By illuminating the MCL-1/BH3-only complexes with 375~nm laser light, we uniformly switched the photoswitchable BH3-only peptides to the \textit{cis}-state and could detect a further increase in $\alpha$-helical content for NOXA, PUMA, and BIM (see Fig.~\ref{fig:CDspec}b), as anticipated from the spacing of 7 amino acids between the two anchoring points of the photoswitch. Previous results on BIM showed that the slight destabilization of the $\alpha$-helix in the \textit{trans}-state is not sufficient to result in unbinding of the peptide in the concentration range needed for IR spectroscopy\cite{Heckmeier2022}. 

\begin{figure*}[t]
	\centering
	\includegraphics[clip, trim=0cm 0cm 0cm 0cm, width=.9\textwidth]{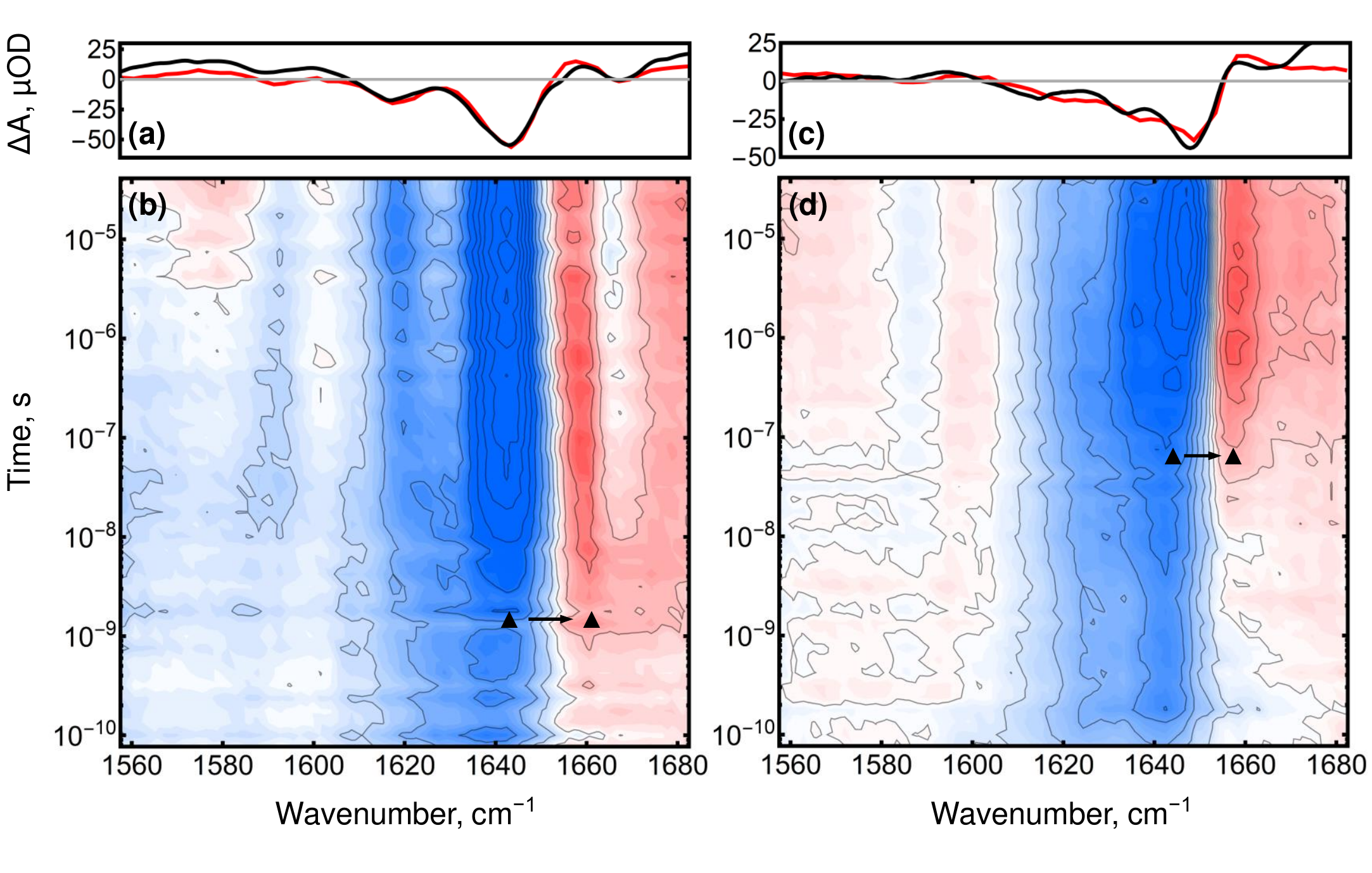}
	\caption{IR spectroscopic analysis of PUMA (a,b) and NOXA (c,d) in complex with MCL-1. (a,c) Steady-state \textit{cis}-to-\textit{trans} difference spectrum (black) and the last kinetic trace at 42 $\mu$s (red). (b,d) Transient \textit{cis}-to-\textit{trans} difference spectra as a function of pump-probe delay time. The triangles mark the blue shift of the amide I band from $\approx$1640~cm$^{-1}$ to $\approx$1660~cm$^{-1}$.} \label{fig:AmideI}
\end{figure*}

For the transient IR experiments, the sample was first prepared in the \textit{cis}-state with the help of a cw-LED at 375~nm. 
Upon subsequent switching of the azobenzene moiety from the \textit{cis}- to the \textit{trans}-state by the irradiation with an ultrashort UV/VIS laser pulse at 420~nm, the abrupt isomerization process perturbs the secondary structure of the peptide,  and in consequence also that of its binding partner MCL-1. To monitor these structural changes, we set our focus on the C=O stretch vibrations of the protein backbone, i.e. the  amide I band  in the region around 1650~cm$^{-1}$. The amide I band serves as an indicator for rearrangements and alterations in the protein structure\cite{barth02}. 
We performed these experiments both in a steady-state manner with the help of a Bruker Tensor~27 FTIR spectrometer (Fig.~\ref{fig:AmideI}a,c), as well as transiently with the pump-probe delay time ranging from picoseconds to 42~$\mu$s  (Fig.~\ref{fig:AmideI}b,d). 
The late time (42~$\mu$s) transient spectra are in essence the same as the steady-state difference spectra (Fig.~\ref{fig:AmideI}a,c red vs black lines), indicating that most of the structural changes have already found their end at this time point.

In order to extract the dynamical processes contained in the transient spectra, we performed global multiexponential fitting, assuming interconverting discrete states with time-invariant spectra\cite{Hobson1998,Kumar2001,Lorenz2006}:
\begin{equation}
d(\omega_i,t_j)=a_0(\omega_i)+\sum_{k} a(\omega_i,\tau_k)e^{-t_j/{\tau_k}}.
\label{LDA}
\end{equation}
Here, we treated the amplitudes $a(\omega_i,\tau_k)$, as well as a common set of time constants $\tau_{k}$ as the free fitting parameters, with the number of exponential terms being restrained to a minimum\cite{VanStokkum2004,Buhrke2020}. 
We fitted the experimental data of PUMA with four states, S$_1$, S$_2$, S$_3$ and a terminal state S$_t$ and three time constants connecting them ($\tau_{12}$=1.6~ns, $\tau_{23}$=18~ns, $\tau_{t}$=1.6~$\mu$s). For NOXA, three states S$_1$, S$_2$, and S$_t$ and two time constants were sufficient ($\tau_{12}$=85~ns and $\tau_{t}$=1.4~$\mu$s). The corresponding time constants in BIM are $\tau_{12}$=9.7~ns,  $\tau_{23}$=150~ns, and $\tau_{t}$=3.6~$\mu$s, see Ref.~\onlinecite{Heckmeier2022}. 

\begin{figure}[t]
	\centering
	\includegraphics[clip, trim=0cm 0cm 0cm 0cm, width=1\textwidth]{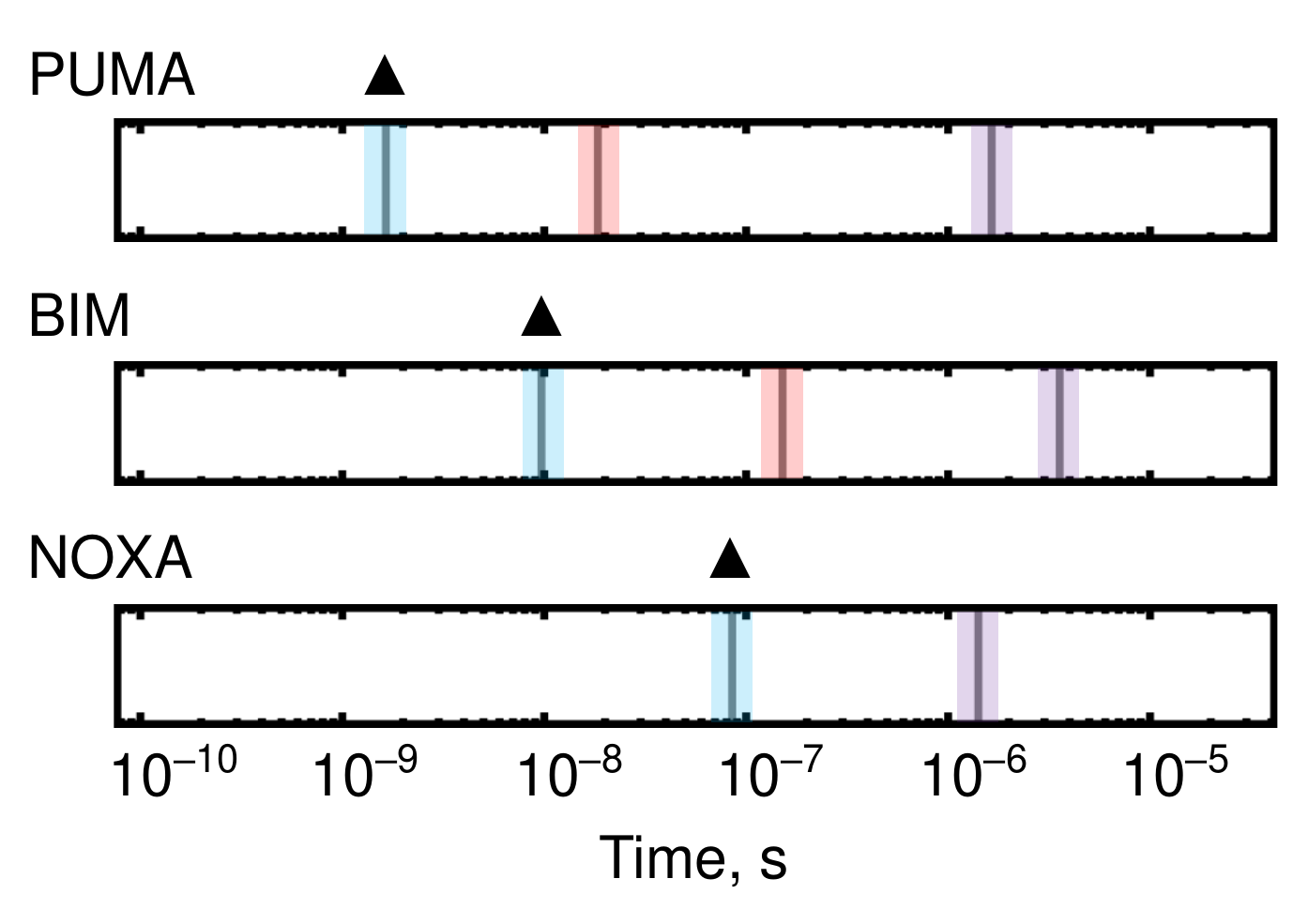}
	\caption{Timescales of dynamical activity for MCL-1/PUMA, MCL-1/BIM, and MCL-1/NOXA upon photo-perturbation. For PUMA and NOXA, the time constants were determined from the transient spectra in Fig.~\ref{fig:AmideI}, while the data for BIM were taken from Heckmeier et al.\cite{Heckmeier2022}. Time constant $\tau_{12}$, associated with the spectral blue shift (and thus marked by triangles), is underlined in blue, time constant $\tau_{23}$ in red, and the terminal time constant $\tau_{t}$ in purple. For NOXA, $\tau_{23}$ cannot be resolved. In Heckmeier et al.\cite{Heckmeier2022}, we reported an additional time constant prior to 100~ps, which however was related to the pump-pulse duration and hence does not reflect a kinetic process.} \label{fig:timescales}
\end{figure} 

The observed timescales are summarized in Fig.~\ref{fig:timescales}. While the last timescale ($\tau_t$) is more or less the same in all three samples, the preceding two processes ($\tau_{12}$ and $\tau_{23}$) vary by almost a factor 100, with PUMA being the fastest and NOXA the slowest. In the case of NOXA, $\tau_{23}$ is not resolved, presumably since it coincides with the terminal process $\tau_{t}$.

\begin{figure}[t]
	\centering
	\includegraphics[clip, trim=1.5cm 0cm 1.5cm 0cm, width=1\textwidth]{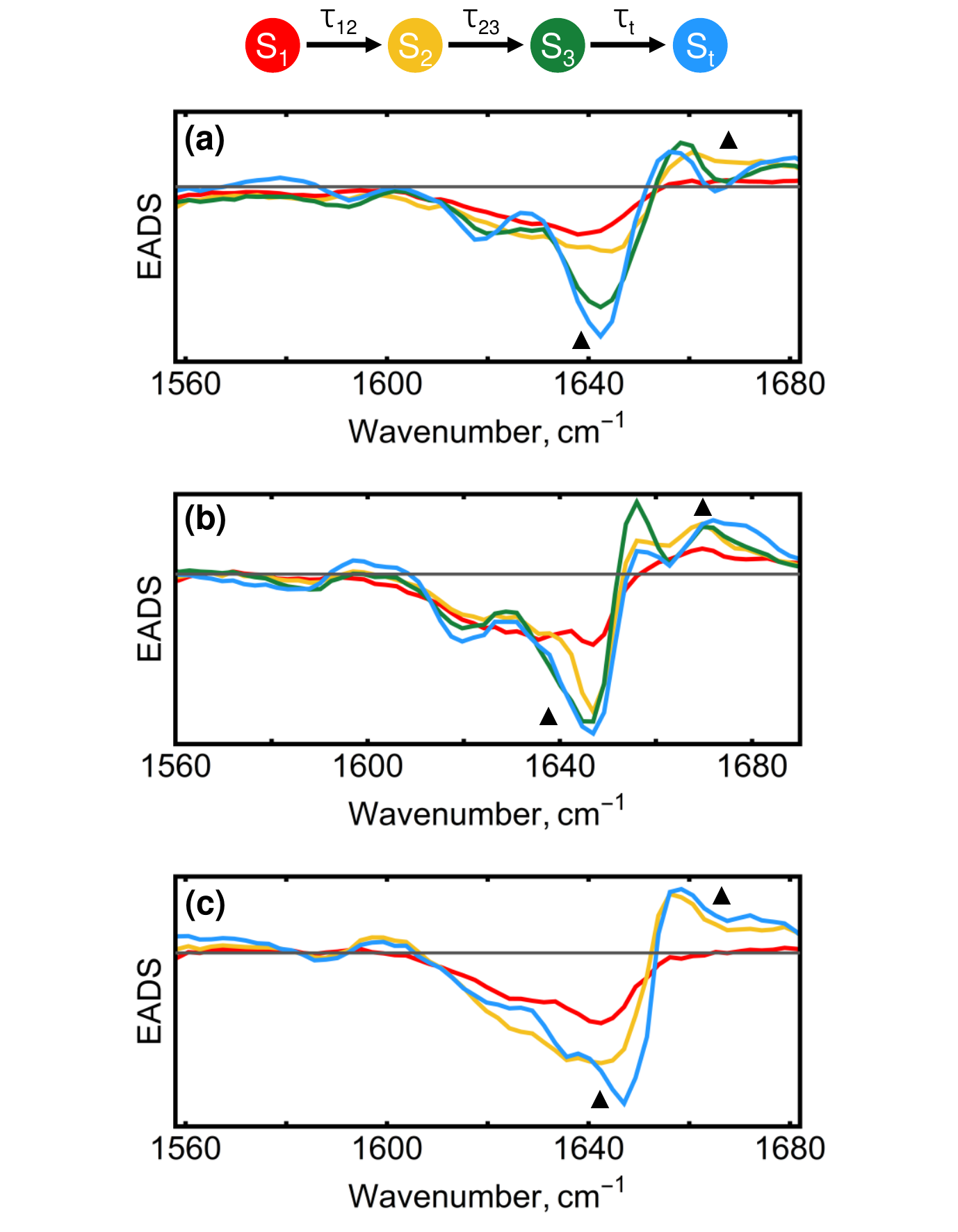}
	\caption{Evolution associated difference spectra (EADS) of photo-perturbed (a) MCL-1/PUMA, (b) MCL-1/BIM (adapted from previous study\cite{Heckmeier2022}), and (c) MCL-1/NOXA. The triangles mark the blue shift of the amide I band from $\approx$1640~cm$^{-1}$ to $\approx$1660~cm$^{-1}$.} \label{fig:EADS}
\end{figure}

Assuming a sequential, unidirectional process (see Fig.~\ref{fig:EADS}, top), we also calculated evolution associated difference spectra (EADS) according to: 
\begin{equation}
d(\omega_i,t_j) = \sum_{k}C_k(t_j)A_k(\omega_i),
\label{separability}
\end{equation}
where $C_k(t_j)$ is the concentration profile of component $k$ as a function of time $t_j$, and $A_k(\omega_i)$ its spectrum at probe frequency $\omega_i$. The spectra  $A_k(\omega_i)$ are a linear combination of the fitting amplitudes $a(\omega_i,\tau_k)$\cite{VanStokkum2004}. 

For all investigated peptides, the first EADS, revealing the response immediately after the pump pulse is over, shows in essence a bleach of the amide~I band (Fig.~\ref{fig:EADS}, red). The subsequent state (Fig.~\ref{fig:EADS}, yellow),  populated with time constant $\tau_{12}$, reveals an additional positive band on the higher-frequency side, and hence a blue-shift of the amide~I band.  This blue shift has been similarly detected for MCL-1/BIM previously\cite{Heckmeier2022} and can be interpreted as a partial $\alpha$-helical unfolding\cite{Barth2007,Huang2002}. In Ref.~\onlinecite{Heckmeier2022}, we had also investigated the response a localized vibrational mode directly associated with the photoswitch, supporting this interpretation. The blue-shift can be identified in the raw data of Fig.~\ref{fig:AmideI} as well, where it is marked with triangles. While the blue-shifted band is observed for all three samples with very similar spectroscopic characteristics, the timescale with which it appears, $\tau_{12}$, varies by almost a factor 100. 

Based on isotope labelling experiments,\cite{Heckmeier2022} the subsequent spectral changes occurring with  $\tau_{23}$ and  $\tau_{t}$ have been attributed to mostly the protein MCL-1 responding to the structural perturbation of its binding partner. We assume the same for PUMA and NOXA.

MCL-1 promiscuously binds to numerous intrinsically disordered inhibitors, BH3-only peptides, at its binding groove \cite{Czabotar2007,Day2008}. MCL-1's centrality in the cancer-related apoptotic networks and its promiscuous nature makes it of high interest for biochemical and pharmaceutical research \cite{Thomas2010,Bolomsky2020}.
In this study, we characterized the promiscuity of MCL-1 on an atomistic level. To that end, we explored the protein response of MCL-1/PUMA and MCL-1/NOXA upon ultrafast photo-perturbation. Together with the recently published data on a photoswitchable MCL-1/BIM variant, we could draw on dynamical information for three of MCL-1's natural inhibitors in a pico- to microsecond time window.

\begin{figure*}[t]
	\centering
	\includegraphics[clip, trim=0cm 0cm 0cm 0cm, width=1\textwidth]{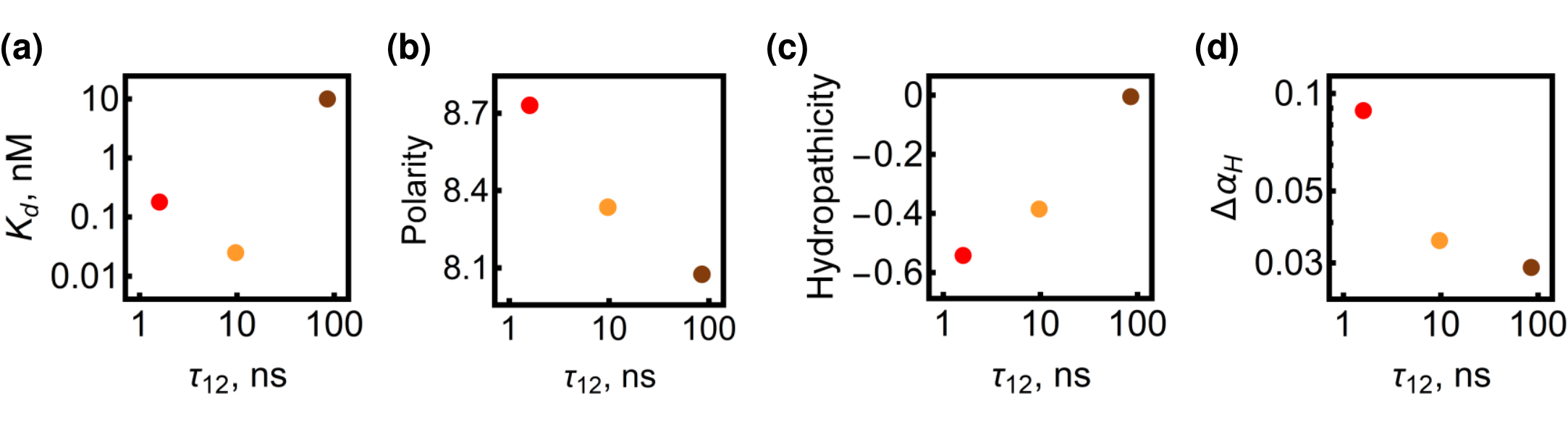}
	\caption{Correlation of (a) the binding affinity of the peptide, (b) its polarity\cite{Grantham1974}, (c) its hydropathicity\cite{Kyte1982}, and (d) the loss in $\alpha$-helicity upon photoswitching (\textit{cis}-to-\textit{trans}) against time constant $\tau_{12}$ for PUMA (red), BIM (orange), and NOXA (brown). The K$_{d}$ values were taken from Dahal et al.,\cite{Dahal2018}, the scores in (b) and (c) were computed from amino acid sequences (see Methods for details), and the data points in (d) from Fig.~\ref{fig:CDspec}.}
	\label{fig:Scores}
\end{figure*}

After the abrupt photo-isomerization of the azobenzene moiety, the three complexes reveal very similar spectroscopic responses, both regarding the late time (steady-state) response (Fig.~\ref{fig:AmideI}a,b), as well as that of the transient intermediates (Fig.~\ref{fig:EADS}).  There is, however, one crucial difference between the three investigated MCL-1/BH3-only complexes: Partial $\alpha$-helical unfolding, occurring with $\tau_{12}$, happens at strongly divergent time points for every BH3-only peptide, see blue lines in Fig.~\ref{fig:timescales}. The fastest response was detected for PUMA ($\tau_{12}$=1.6~ns), succeeded by BIM ($\tau_{12}$=9.7~ns), and finally by NOXA ($\tau_{12}$=85~ns). 
Interestingly, the time point of partial $\alpha$-helical unfolding does not correlate with the binding affinity of the peptide\cite{Dahal2018}, but significantly with computed scores for its polarity\cite{Grantham1974} and hydropathicity\cite{Kyte1982}, as well as with the decline of $\alpha$-helical content due to photoswitching (Fig.~\ref{fig:Scores}). The stronger the decrease in $\alpha$-helicity is upon \textit{cis}-to-\textit{trans} isomerization, the faster is its response. The same is true for increasing hydrophilicity as well as polarity.    

Disrupting PUMA, which reveals the fastest response, with the \textit{trans}-state azobenzene moiety may result in unfavorable contacts between hydrophilic side chains of PUMA and the hydrophobic interface of the binding groove of MCL-1. This strong tension could be dissolved by a rapid partial $\alpha$-helical unfolding. Interestingly, a computational study solely on PUMA highlighted the increased $\alpha$-helical propensity at the C-terminal relative to the rest of the peptide\cite{Chebaro2015}, the same region that we destabilize in our experiments. Apparently, the perturbation of this region which is innately more likely to form ordered structures, leads to conformational tension and thus to the fast protein response.
In contrast, the slow opponent NOXA -- inherently more hydrophobic and less polar -- is not confronted with comparable tensions, thus it is not forced to rearrange quickly and thus unfolds later. The discrepancies between the peptides may also arise from breaking peptide-exclusive electrostatic interactions at the putative contact interface (Fig.~\ref{fig:Peptides}, yellow residues) upon photoswitching. Electrostatic interactions at the contact interface are known to support the complex formation of MCL-1 and BH3-only peptides, namely those formed by MCL-1~Lys234 and Glu9 of NOXA or PUMA (numbering according to the alignment in Fig.~\ref{fig:Peptides})\cite{Czabotar2007,Chu2017}, MCL-1~Asp256/peptide~Arg15\cite{Stewart2010,Chu2017}, and MCL-1 Arg263/peptide~Asp19\cite{Stewart2010,Chu2017}.

These observations pose the question of how the peptide-specific differences correlate with the promiscuous nature of the MCL-1/BH3-only complexes on a cellular level. Pro-apoptotic BH3-only proteins are ``damage sensors'' and abundantly expressed when cells suffer cytotoxic stress \cite{Adams2007,Czabotar2014,Roufayel2022}. Although their binding domains share the same $\alpha$-helical structure of similar length and identical hydrophobic heptad pattern, as well as the same binding site at MCL-1, BH3-only proteins differ substantially in their relationship inside the BCL-2 family and their interaction pattern (Fig.~\ref{fig:Introduction}). BIM and PUMA bind the full spectrum of the anti-apoptotic BCL-2 family, and also activate the effector proteins BAK and BAX, whereas NOXA selectively inhibits MCL-1 and BCL-2, another member of the same protein family\cite{Kale2018}. BIM and PUMA stabilize MCL-1, while NOXA promotes its degradation \cite{Czabotar2007,Willis2005}. 

Studies with truncated or mutated versions of NOXA demonstrated that the C-terminal region of the binding domain regulates the stability of the MCL-1/NOXA complex and therefore is important to control MCL-1/NOXA degradation\cite{Willis2005,Czabotar2007,Pang2014} (Fig.~\ref{fig:Introduction}, left). For its biological function, this region of NOXA has to bind to MCL-1, which requires a certain ``structural resilience'' for this region, even when the whole complex is destabilized in the proteolytic process. We delimit this ``structural resilience'' from the already established terms ``stability''\cite{Murphy1992} and ``structural plasticity''\cite{Matthews2013}. The former is classically used in a thermodynamic context to describe how partners in a protein complex form and maintain folded conformations  \cite{Murphy1992,Pace1996,Guerois2002,Majewski2019}. The latter is frequently used to characterize the ability of promiscuous proteins, e.g. BCL-2-type\cite{Smits2008}, chaperones\cite{Tapley2009,Ashcroft2002}, Trypsin\cite{Plattner2015}, or protein kinases \cite{Huse2002}, to flexibly bind various, different binding partners at the same interface. With ``structural resilience'', we have in mind a kinetic stabilisation, to contrast to a thermodynamics one.

Our results show that even after photo-induced destabilization, the photo-perturbed C-terminus of the NOXA peptide remains folded more than 10 times longer than the equivalent region in other peptides (Fig.~\ref{fig:timescales}). While that is definitely speculative, the high structural resilience of NOXA's C-terminus inside the binding pocket of MCL-1 could help to remain in place, even when the whole complex is confronted with major rearrangements leading to the potent proteolysis of MCL-1. Future experiments could test this hypothesis by connecting a mutational analysis of NOXA and \textit{in vivo} screening of MCL-1 proteolysis with monitoring the peptide's structural resilience by IR spectroscopy.

Different to NOXA, PUMA and BIM increase the stability of MCL-1\cite{Mei2005,Czabotar2007} and block the binding groove of MCL-1 (Fig.~\ref{fig:Introduction}, right). This in turn limits MCL-1's ability to bind the effector proteins BAK and BAX. The competition between BIM, PUMA, and the effector proteins for MCL-1 manifests in the extremely high affinities of BIM (K$_{d}$=25 pM) and PUMA (K$_{d}$=180 pM) in comparison to the already high affinities for the effector proteins BAK (K$_{d}$=1.4~nM) and BAX (K$_{d}$=22~nM)\cite{Dahal2018}. 
Transient IR spectroscopy demonstrated that the structural resilience of PUMA and BIM is smaller than that of NOXA, presumably because their natural role is different, binding the partner in a highly-stable complex with little structural flexibility. Any disturbance seems to result in a fast adaptation, i.e. the partial unfolding that we observed in our experiments. 

In summary, our study reveals insights into the promiscuity of the anti-apoptotic MCL-1 for the intrinsically disordered binding domains of BH3-only proteins. By using transient IR spectroscopy in combination with photo-switchable protein complexes, we quantified the dynamic response of BIM, PUMA, and NOXA in the  binding pocket of MCL-1 upon photo-perturbation. All peptides show partial $\alpha$-helical unfolding, however on very different time-scales.  The correlations in Fig.~\ref{fig:Scores} are indicating that the speed of protein response in our system is coupled to the extent of structural rearrangements (Fig.~\ref{fig:Scores}d) and possibly linked to favourable or unfavourable interactions, caused by the perturbation. Computational studies could test this hypothesis.

NOXA is structurally more resilient than BIM and PUMA. This finding reveals a new viewpoint on the nature of BH3-only peptides, which could help to better understand promiscuous protein-protein interactions in general, as well as the design of novel molecules and peptides to therapeutically manipulate the oncologically relevant MCL-1/BH3-only complex. 

\section*{Materials and Methods}

\subsection{Peptide preparation}
The BH3 domains of PUMA (EEQWAREIGAQLRCMADDLNCQYER) and NOXA (RAELPPEFAAQLRCIGDKVYCTWSAP), both containing cysteine mutations with a spacing of 7 amino acids, were synthesized using solid state peptide synthesis on a Liberty 1 peptide synthesizer (CEM corporation, Matthews, NC, USA). These peptides were purified analogously to the BIM variant (GGSGRPEIWIAQELRCIGDEFNCYYARRV), which was investigated in a preceding study \cite{Heckmeier2022}. The watersoluble photoswitch 3,3'-bis(sulfonato)-4,4'-bis(chloroacetamido)azobenzene (BSBCA) was subsequently covalently linked to the cysteine residues, as described before \cite{Zhang2003,Jankovic2019}. The successful linkage, purity, as well as the integrity of the peptide was controlled via mass spectrometry.

\subsection{Protein preparation}
Human MCL-1 (hMCL-1$\Delta$N-$\Delta$C, 171-327, C286S\cite{Liu2014}) was expressed in \textit{Escherichia coli} BL21. The cells were grown until they reached OD$_{600}$=0.6 and induced with 0.75 mM isopropyl $\beta$-D-1-thiogalactopyranoside. After the induction, the cells were incubated for 20 hours at 30° C until they were harvested end lysed using sonication. The proteins were purified under native conditions via Ni-affinity chromatography and a His$_6$-Tag at the N-terminus of the protein. The N-terminal His$_6$-Tag was removed by 3C protease cleavage. All experiments were performed in 50~mM Tris (pH~8) and 125~mM NaCl. The integrity, as well as the purity of the protein sample was controlled via mass spectrometry. For IR spectroscopy, H$_2$O containing sample buffer was exchanged against D$_2$O based buffer via spin column centrifugation. To circumvent the contamination with atmospheric H$_2$O, the sample was kept in a water-vapor free nitrogen environment.

\subsection{Transient IR spectroscopy}
For pump-probe measurements, we used two electronically synchronized 2.5 kHz Ti:sapphire oscillator/regenerative amplifier femtosecond laser systems (Spectra Physics), allowing a delay of maximally 45~$\mu$s \cite{Bredenbeck2004}. For the pump pulses, one laser system was tuned to 840~nm and then brought to 420~nm pulses via second harmonic generation in a $\beta$-BaBO$_{4}$ crystal, later needed for \textit{cis}-to-\textit{trans}-isomerization of the photoswitch. After light amplification, the compressor has been bypassed, resulting in $\approx$60~ps stretched pulses, in order to reduce sample deposition on the sample cell windows. At the sample cell, the power was 3~$\mu$J per pulse, focused to a $\approx$140 $\mu$m beam diameter. The second laser system was used to generate midIR probe pulses in an optical parametric amplifier (100~fs, spot size 110~$\mu$m, center wavenumber 1625~cm$^{-1}$)\cite{Hamm2000}. To prepare the investigated samples for the spectroscopic experiments, the crosslinked peptides and MCL-1 were mixed in an 1:1 ratio in D$_2$O with a total complex concentration of 1~mM. The protein samples were constantly circulated in a closed-cycle flow cell system, comprising a CaF$_{2}$ measurement cell (50~$\mu$m optical path length) and a reservoir. The investigated sample was irradiated with a 375~nm continuous wave diode laser (90~mW, CrystaLaser) before entering the measurement cell, in order to prepare $>$85\% of the sample in \textit{cis}-state. 

\subsection{Peptide parameter computation}
To better understand the differences of the various BH3-only peptides, we calculated parameters  for the polarity according to Grantham,\cite{Grantham1974} and the hydrophobicity/hydropathicity according to Kyte \& Doolitle\cite{Kyte1982} from the sequences of PUMA, BIM, and NOXA, using the Expasy/ProtScale server (http://web.expasy.org/protscale/).  The calculations were executed with a window size of 7~residues (interval length for the computation) and the same weight for every residue in that interval (relative weight = 100$\%$)\cite{Gasteiger2005}.

\section*{Abbreviations}
BH, BCL-2 homology; BSBCA, 3,3'-bis(sulfonato)-4,4'-bis-(chloroacetamido) azobenzene; CD, Circular Dichroism; EADS, evolution associated difference spectra; IR, Infrared; MCL-1, Myeloid Cell Leukemia~1.  

\section*{Author Information}
\noindent \textbf{Corresponding Author} \\
* E-mail: philipp.heckmeier@chem.uzh.ch\\
\\
\textbf{ORCID identifiers} \\[0.25cm]
\begin{tabular}{ l l }
	Philipp J. Heckmeier       	& \quad 0000-0002-7311-977X \\
	Jeannette Ruf               & \quad 0000-0003-0495-6459 \\
	Brankica G. Jankovi\'{c}    & \quad 0000-0002-8073-3014 \\
	Peter Hamm                 	& \quad 0000-0003-1106-6032 \\
\end{tabular}
\\[0.25cm]
\textbf{Notes} \\
The authors declare no competing financial interest.\\

\noindent\textbf{Acknowledgement}
We thank Markus B. Glutz for the synthesis of the peptides and Functional Genomics Center Zurich for their work on the mass spectrometry analysis. We thank Roland Zehnder for the technical support and Kerstin T. Oppelt, Gerhard Stock, and Steffen Wolf for fruitful discussions. The work has been supported by the Swiss National Science Foundation (SNF) through Grant 200020B\_188694/1.\\

\noindent\textbf{Data Availability:} The data that support the findings of this study are openly
available in Zenodo (the link will be provided at the proof stage.)\\



\makeatletter
\def\@biblabel#1{(#1)}
\makeatother

\def\bibsection{\section*{}} 

\noindent\textbf{References:}
\vspace{-1.5cm}

\providecommand{\latin}[1]{#1}
\makeatletter
\providecommand{\doi}
  {\begingroup\let\do\@makeother\dospecials
  \catcode`\{=1 \catcode`\}=2 \doi@aux}
\providecommand{\doi@aux}[1]{\endgroup\texttt{#1}}
\makeatother
\providecommand*\mcitethebibliography{\thebibliography}
\csname @ifundefined\endcsname{endmcitethebibliography}
  {\let\endmcitethebibliography\endthebibliography}{}


\end{document}